\documentclass[t11pt,superscriptaddress,reprint,preprint,preprintnumbers]{revtex4}

\usepackage{braket}
\usepackage{amsmath,amssymb,latexsym, amsthm}
\usepackage{tikz}
\usepackage{textcomp}
\usepackage{eucal}

\usepackage{booktabs}
\usepackage[framemethod=tikz,xcolor=true]{mdframed}

\DeclareMathOperator{\Corr}{Corr}
\DeclareMathOperator{\Cov}{Cov}
\DeclareMathOperator{\std}{std}
\DeclareMathOperator{\E}{\mathbb{E}}
\DeclareMathOperator{\var}{Var}

\newtheorem*{theorem}{Theorem}

\begin{document}

\title{For Fixed Control Parameters the Quantum Approximate Optimization Algorithm\textquotesingle s Objective Function Value Concentrates for Typical Instances}

\author{Fernando G.S.L. Brand\~ao}
\affiliation{Google Inc., Venice, California 90291, USA
}
\affiliation{Institute for Quantum Information and Matter, California Institute of Technology, Pasadena, California 91125, USA
}

\author{Michael Broughton}
\affiliation{Department of Computer Science, University of Waterloo, Waterloo, Ontario, N2L 3G1, Canada
}
\affiliation{Google Inc., Venice, California 90291, USA
}
\author{Edward Farhi}
\affiliation{Google Inc., Venice, California 90291, USA
}
\affiliation{Center for Theoretical Physics, Massachusetts Institute of Technology, Cambridge, MA 02139
}
\author{Sam Gutmann, and Hartmut Neven}

\affiliation{Google Inc., Venice, California 90291, USA
}


\date{\today}

\begin{abstract} 

The Quantum Approximate Optimization Algorithm, QAOA, uses a shallow depth quantum circuit to produce a parameter dependent state. For a given combinatorial optimization problem instance, the quantum expectation of the associated cost function is the parameter dependent objective function of the QAOA.  We demonstrate that if the parameters are fixed and the instance comes from a reasonable distribution then the objective function value is concentrated in the sense that typical instances have (nearly) the same value of the objective function. This applies not just for optimal parameters as the whole landscape is instance independent. We can prove this is true for low depth quantum circuits for instances of MaxCut on large 3-regular graphs. Our results generalize beyond this example. We support the arguments with numerical examples that show remarkable concentration. For higher depth circuits the numerics also show concentration and we argue for this using the Law of Large Numbers. We also observe by simulation that if we find parameters which result in good performance at say 10 bits these same parameters result in good performance at say 24 bits. These findings suggest ways to run the QAOA that reduce or eliminate the use of the outer loop optimization and may allow us to find good solutions with fewer calls to the quantum computer.
\end{abstract}

\maketitle

\section{Introduction}

Soon we will have gate model quantum computers that can run shallow depth quantum circuits on scores of qubits. Even without perfect fidelity we will learn about quantum algorithms in a regime where classical simulation is no longer practical.  A good candidate to run on near term devices, as well as on larger devices with high fidelity, is the Quantum Approximate Optimization Algorithm or QAOA \cite{farhi-2014}.  For certain combinatorial search problems, the shallowest depth version of QAOA has worst case performance guarantees that beat random guessing \cite{farhi-2014, farhi-2015}, but not the best classical algorithms for these problems \cite{halperin-2004,Boaz-2015,hastad-2015}.  At higher depth we do not know if the QAOA will outperform classical algorithms. This question needs to be explored analytically, through simulation, and by running on actual hardware.  In all cases we need strategies for picking parameters that optimize performance. For a given instance of a combinatorial search problem one strategy is to seek to optimize the objective function by varying the control parameters attempting to go uphill to find good solutions.  In this paper we look at instances of combinatorial search problems that are chosen randomly from some fixed instance distribution. We show that the parameter landscape of the objective function is (nearly) independent of the chosen instance.  This points to a strategy for finding good parameters. Take one instance of the problem and work hard to get good parameters.  This may be computationally expensive. But once this has been done, these same parameter values will yield good values of the cost function on other randomly chosen instances. In other words the amortized cost of solving instances goes to zero inversely with the number of instances being studied. Recent work has also looked at strategies for picking optimal parameters that reduce calls to the quantum computer \cite{zhou-2018,crooks-2018,hastings-2016}, while others have also remarked on the independence of the function value on the instance \cite{zhou-2018, bravyi-2018}.

We begin by reviewing the QAOA and setting notation. The goal is to find a good approximation ratio for a combinatorial search problem over $n$ bits. We denote bit strings as $z = z_1 z_2 \dots z_n$ and the search problem is specified by $m$ clauses each of which is defined on a subset of the bits.  Associated with each clause $\alpha$ is a cost function $C_{\alpha}(z)$ which is 1 if $z$ satisfies the clause $\alpha$ and is 0 if $z$ does not.  The total cost function is then
\begin{equation}\label{cost_eqn}
    C(z) = \sum_{\alpha = 1}^{m} C_{\alpha}(z) \text{ .}
\end{equation}
Let $C_{\text{max}}$ be the maximum over all $z$ of $C(z)$.  If an algorithm proposes $z^{*}$ as a candidate solution then the approximation ratio is
\begin{equation}\label{a_ratio}
    A = \frac{C(z^{*})}{C_{\text{max}}}
\end{equation}
and the goal of any algorithm is to make $A$ as close to 1 as possible. If each clause is satisfied on half of its input values then random guessing gives $A=(m/2)/C_{\text{max}}$ which is lower bounded by $1/2$.

The quantum computer operates in a $2^n$ dimensional Hilbert space with a computational basis $\ket{z}$.  Now the quantum operator $C$ is defined as
\begin{equation}
    C \ket{z} = C(z) \ket{z} \text{.}
\end{equation}
We introduce a unitary operator that depends on $C$ and a parameter $\gamma$ as
\begin{equation}\label{driver}
    U(C, \gamma) = e^{-i \gamma C} = \prod_{\alpha=1}^{m} e^{-i \gamma C_{\alpha}} \text{ .}
\end{equation}
Note that each term in the product commutes with the others and because $C$ is integer valued, $\gamma$ is an angle between $0$ and $2 \pi$.  We introduce another operator
\begin{equation}
    B = \sum_{j=1}^{n} X_{j}
\end{equation}
where $X_{j}$ is the Pauli $X$ operator on qubit $j$, and an associated unitary that depends on a parameter $\beta$
\begin{equation}\label{mixer}
    U(B, \beta) = e^{-i \beta B} = \prod_{j=1}^{n} e^{-i \beta X_j}
\end{equation}
where $\beta$ is an angle between $0$ and $\pi$. 

The QAOA consists of an alternation of operators of the form (\ref{driver}) with operators of the form (\ref{mixer}) acting on the initial state
\begin{equation}\label{uniform_sup}
    \ket{s} = \frac{1}{\sqrt{2^n}} \sum_{z} \ket{z} \text{.}
\end{equation}
Each layer has its own parameters. Let $\boldsymbol{\gamma} = \gamma_{1} \dots \gamma_{p}$ and $\boldsymbol{\beta} = \beta_{1} \dots \beta_{p}$ and define the quantum state
\begin{equation}\label{alternation}
    \ket{\boldsymbol{\gamma}, \boldsymbol{\beta}} = U(B, \beta_{p}) U(C, \gamma_{p}) \dots U(B, \beta_{1}) U(C, \gamma_{1}) \ket{s} \text{.}
\end{equation}
For a given instance of a problem, the associated objective function on $2p$ parameters is
\begin{equation}\label{objective_f}
    F_{p}(\boldsymbol{\gamma}, \boldsymbol{\beta}) = \bra{\boldsymbol{\gamma}, \boldsymbol{\beta}} C \ket{\boldsymbol{\gamma}, \boldsymbol{\beta}} \text{.}
\end{equation}
We use `cost' to refer to Eq. (\ref{cost_eqn}) which depends on strings and `objective' to refer to Eq. (\ref{objective_f}) which depends on parameters. For a fixed instance and value of $p$, we view (\ref{objective_f}) as the parameter landscape. The ultimate goal is to find the high points in the landscape. For now we want to see how the landscape varies as we look at different instances which come from some instance distribution. For example we might look at $n$ bit instances of Max3Sat where the distribution is uniform over all instances with the ratio of the number of clauses to the number of bits fixed. Or we could look at MaxCut on graphs where each edge is included with probability $3/(n-1)$ so the expected value of the valence is 3.  However in this paper we focus on MaxCut where the distribution is over all 3-regular graphs. Our results will generalize beyond the examples but they contain the essence of the more general arguments.

\section{Fixed $\protect\scalebox{1.25}{$\mathit{p}$}$ with $\protect\scalebox{1.25}{$\mathit{n}$}$ large}\label{fixed_p_section}

In this section we consider how the objective function varies when we pick typical instances from a given distribution while holding $p$ fixed and taking $n$ to be large. We start with MaxCut whose cost function can be written as
\begin{equation}\label{maxcut_cost}
    C = \sum_{\langle jk \rangle} C_{\langle jk \rangle}
\end{equation}
and
\begin{equation}\label{maxcut_cost2}
    C_{\langle jk \rangle} = \frac{1}{2}(-Z_{j} Z_{k} + 1) \text{.}
\end{equation}
Now the objective function can be written as
\begin{equation}\label{objective_big}
\begin{split}
    F_{p}(\boldsymbol{\gamma}, \boldsymbol{\beta}) &= \sum_{\langle jk \rangle} \bra{s} U^{\dag}(C, \gamma_{1}) \dots U^{\dag}(B, \beta_p) C_{\langle jk \rangle} U(B, \beta_{p}) \dots U(C, \gamma_{1}) \ket{s} \\
                         &= \sum_{\langle jk \rangle} F_{\langle jk \rangle} (\boldsymbol{\gamma}, \boldsymbol{\beta})
\end{split}
\end{equation}
where we will make use of the fact that the objective function is a sum over individual edge functions. We first consider only 3-regular graphs and the shallowest depth for the QAOA which is $p=1$.  Consider one term in equation (\ref{objective_big}) that comes from the edge $\langle jk \rangle$.  We can write the relevant piece as
\begin{equation}
    F_{\langle jk \rangle}(\gamma_{1}, \beta_{1}) = \bra{s} U^{\dag} (C, \gamma_{1}) U^{\dag} (B, \beta_{1}) Z_{j} Z_{k} U(B, \beta_{1}) U(C, \gamma_{1}) \ket{s} \text{.}
\end{equation}
The effects of the conjugation by $U(B,\beta_{1})$ is to rotate $Z_{j}$ and $Z_{k}$ about the x-axis so $Z_{j} \rightarrow Z_{j} \cos(2 \beta_{1}) + Y_{j} \sin(2\beta_{1})$ and similarly for $Z_{k}$.  Still only qubits $j$ and $k$ are involved.  Now conjugation by $U(C,\gamma_{1})$ introduces only the qubits that are immediately connected to qubits $j$ and $k$ on the 3-regular graph. There are 3 possibilities illustrated here
\begin{equation}\label{subgraph_types}
    \includegraphics[scale=.95,trim=50 40 0 0]{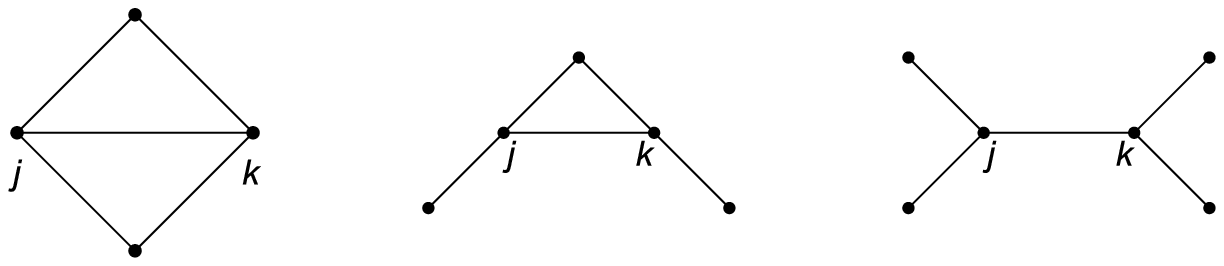} \text{ \Large$ .$}
\end{equation}\\
So in the sum over all edges there are only 3 functions that can contribute
\newcommand{\squaresubgraph}[1]{%
\begin{tikzpicture}[#1]%
\draw (0ex,0ex) -- (1ex,0ex);%
\draw (0ex,0ex) -- (0.5ex, 0.5ex);
\draw (0.5ex, 0.5ex) -- (1ex,0ex);
\draw (0ex, 0ex) -- (0.5ex, -0.5ex);
\draw (0.5ex, -0.5ex) -- (1ex,0ex);
\filldraw [black] (0ex, 0ex) circle (0.1pt);
\filldraw [black] (1ex, 0ex) circle (0.1pt);
\filldraw [black] (0.5ex, 0.5ex) circle (0.1pt);
\filldraw [black] (0.5ex, -0.5ex) circle (0.1pt);
\end{tikzpicture}%
}
\newcommand{\trianglesubgraph}[1]{%
\begin{tikzpicture}[#1]%
\draw (0ex,0ex) -- (1ex,0ex);%
\draw (0ex,0ex) -- (0.5ex, 0.5ex);
\draw (0.5ex, 0.5ex) -- (1ex,0ex);
\draw (0ex, 0ex) -- (-0.5ex, -0.5ex);
\draw (1ex, 0ex) -- (1.5ex,-0.5ex);
\filldraw [black] (0ex, 0ex) circle (0.1pt);
\filldraw [black] (1ex, 0ex) circle (0.1pt);
\filldraw [black] (0.5ex, 0.5ex) circle (0.1pt);
\filldraw [black] (-0.5ex, -0.5ex) circle (0.1pt);
\filldraw [black] (1.5ex,-0.5ex) circle (0.1pt);
\end{tikzpicture}%
}
\newcommand{\treesubgraph}[1]{%
\begin{tikzpicture}[#1]%
\draw (0ex,0ex) -- (1ex,0ex);%
\draw (0ex,0ex) -- (-0.5ex, 0.5ex);
\draw (1ex,0ex) -- (1.5ex, 0.5ex);
\draw (0ex, 0ex) -- (-0.5ex, -0.5ex);
\draw (1ex, 0ex) -- (1.5ex,-0.5ex);
\filldraw [black] (0ex, 0ex) circle (0.1pt);
\filldraw [black] (1ex, 0ex) circle (0.1pt);
\filldraw [black] (1.5ex, 0.5ex) circle (0.1pt);
\filldraw [black] (-0.5ex, 0.5ex) circle (0.1pt);
\filldraw [black] (-0.5ex, -0.5ex) circle (0.1pt);
\filldraw [black] (1.5ex,-0.5ex) circle (0.1pt);
\end{tikzpicture}%
}
\begin{equation}\label{weightedpictures}
    F_{1}(\gamma_{1}, \beta_{1}) = w_{\squaresubgraph{scale=1.5}} F_{\squaresubgraph{scale=1.5}}(\gamma_{1}, \beta_{1}) + w_{\trianglesubgraph{scale=1.5}} F_{\trianglesubgraph{scale=1.5}}(\gamma_{1}, \beta_{1}) + w_{\treesubgraph{scale=1.5}} F_{\treesubgraph{scale=1.5}}(\gamma_{1}, \beta_{1})
\end{equation}
where $w_{\squaresubgraph{}}$ is the number of edges that locally look like the first picture in Eq. (\ref{subgraph_types}) and similarly for $w_{\trianglesubgraph{}}$ and $w_{\treesubgraph{}}$.  Now we introduce fractions of each type so that $f_{\squaresubgraph{}} =  w_{\squaresubgraph{}} /m$, $f_{\trianglesubgraph{}} = w_{\trianglesubgraph{}} / m$ and $f_{\treesubgraph{}}= w_{\treesubgraph{}}/ m$ where $m = 3n/2$ is the number of edges.  So now
\begin{equation}
    F_{1}(\gamma_{1}, \beta_{1}) = \Big[f_{\squaresubgraph{}} F_{\squaresubgraph{}} (\gamma_{1}, \beta_{1}) +
                                    f_{\trianglesubgraph{}} F_{\trianglesubgraph{}} (\gamma_{1}, \beta_{1}) + 
                                    f_{\treesubgraph{}} F_{\treesubgraph{}} (\gamma_{1}, \beta_{1}) \Big] \cdot m \text{ .}
\end{equation}
The key point is that for the distribution at hand, as $n$ grows, the fractions $f_{\squaresubgraph{}}$, $f_{\trianglesubgraph{}}$ and $f_{\treesubgraph{}}$ concentrate. In this particular case as $n$ gets big,  for a typical 3-regular graph, almost all edges' neighborhoods locally look like trees. So $f_{\treesubgraph{}}$ goes to 1 and the other two are of order $1/n$. Regardless of the parameter values, the objective function is the same up to order $1/n$ for almost all 3-regular graphs. 

Consider what happens as we increase $p$ still working with 3-regular graphs. Look at edge $\langle jk \rangle$. The successive conjugations by the cost function unitary bring into play qubits that are at most $p$ away from $j$ and $k$. For fixed $p$ the number of subgraph types is fixed and is independent of the number of bits.  As $n$ gets large, on random 3-regular graphs, the subgraph fractions will concentrate. (For a random 3-regular graph, in fact there is one dominant subgraph type, a tree.) This means that for fixed $p$ as $n$ gets large, for fixed parameters, the objective function value doesn't depend on which $n$-bit-3-regular graph is randomly chosen. See also \cite{zhou-2018}.

Consider now a different distribution of graphs, for example where each edge is included with probability $3/(n-1)$ so the average valence is 3.  With this distribution, tossed graphs have isolated vertices and disconnected components.  So we might focus on the largest connected component.  Now at $p=1$ there are more subgraph types than we found on 3-regular graphs. Coming out of each of the two vertices on edge $\langle jk \rangle$ can be $0, 1, 2 , 3 \dots$ edges. Still the fraction of each subgraph type will concentrate as $n$ gets large, dominated by trees.  Since there is now a family of trees, there will be $\sqrt{n}$ fluctuations on the number of tree types and so the total cost function value will concentrate with $\sqrt{n}$ fluctuations.  With $p$ larger than 1 but fixed as $n$ grows, there will still be concentration in the number of subgraph types and correspondingly concentration in the total cost function.

For $p$ fixed and $n$ large, with  fixed parameters $\boldsymbol{\gamma}$ and $\boldsymbol{\beta}$ we have shown that the cost function will be the same on all typical instances of MaxCut on 3-regular graphs or Renyi graphs with probability $3/(n-1)$.  The same reasoning will apply to other combinatorial search problems which have the restriction that the number of clauses in which any variable can appear does not grow with $n$ or at least grows only slowly with high probability.  Otherwise for fixed $p$, the relevant subgraph types will cover multiple times the instance graph and our fixed $p$ arguments will not apply.

There are well known theorems about well-behaved functions (e.g. with bounded Lipschitz constants) of strings which state that the function value concentrates when the strings are chosen randomly. For example consider a specialized version of the McDiarmind/Hoeffding/Azuma inequality:

\begin{theorem}
Let bit strings $x_{1},\dots, x_{L}$ be generated as independent random variables according to some distribution.  Consider an arbitrary function of the strings $f(x_{1},\dots, x_{L})$ which has the property that if any bit is flipped the function value changes by no more than a constant $c$. Then 
\begin{equation}\label{hoeffding_exp}
    \Pr \Big(\Big\lvert f(x_{1},\dots, x_{L}) - \E[f] \Big\rvert > t \Big) < \exp \bigg(\frac{-2t^2}{Lc^2} \bigg) \text{ .}
\end{equation}
\end{theorem}

An $n$ vertex graph has $\binom{n}{2}$ possible edges so we can think of a graph as described by $\binom{n}{2}$ bits where each bit says if an edge is present or not. So now think of the function $F$ in (\ref{objective_f}) as a function of $\binom{n}{2}$ bits. We can show that the function (\ref{objective_f}) does not change much when an edge is removed if $p$ is fixed, as long as the maximum valence does not grow with $n$. However now $L$ is of order $n^2$ so for the right hand side of (\ref{hoeffding_exp}) to be small we need $t$ of order $n$ and we are not getting good concentration.  But what we have done is too simple.  The theorem above does not notice that we are working with say Renyi graphs with edge probability $3/(n-1)$. There is a graph version of this theorem (see Remark 3.25 of \cite{Bordenave16}) which gives a concentration result for the Erdos-Renyi graph case.  But in our attempts to use it we found concentration that was not as tight as what we obtained with the subgraph argument.

However there is an advantage to this approach. Suppose we consider a generalization of the QAOA to situations where for each $p$ we have angles $\boldsymbol{\gamma}$ for each clause. Or consider weighted MaxCut. Now our subgraph counting arguments will not work because each edge is unique and we can not count subgraph types. Still the graph concentration techniques can be used. But if $p$ grows with $n$, we do not know how to meet the hypothesis of the theorems which is that the function value (\ref{objective_f}) does not change much if we remove an edge. However in the large $p$ case we have other arguments which will be presented in section \ref{higher_p_section}.

\section{Numerics that Show Concentration}
Here we present some numerics that support the arguments made in the previous section and prepare for the next section on higher $p$.  We look at the MaxCut cost function working at 20 bits tossing 3-regular graphs.  The number of edges in each graph is 30 and to decrease some fluctuations in some parameter regimes we put the further restriction that each graph has a MaxCut value of 26, which means that the largest possible value of the cost function is the same for all graphs.

If we select the control parameters at random, especially at high $p$, one could argue that we are producing a random state and the quantum expectation of any clause would be near 1/2, so the total objective function would be near 15.  In this sense the fact that we get the same value of the objective function regardless of which random graph is tossed would have an alternate explanation based on random states. So for our numerics we work with three regimes for each value of $p$. For the first we select values of the parameters so that the objective function has a small value, much less than 15. In section \ref{from_small_to_large} we will explain how these parameters are chosen. But for each $p$, once the $2p$ parameters are fixed we toss 25 random graphs and evaluate the quantum expectation of the cost function. The first column of Table 1 shows the results for $p$ ranging from 2 to 7. We give the average over the 25 sample graphs and the standard deviation of the 25 samples. The concentration is evident. In the next column for each p we select $2p$ angles at random.  Typically the quantum expectation of the cost function is near 15.  Again we toss 25 random 3-regular graphs and report the average over the 25 and the standard deviation.  For the data in the third column for each $p$ we select $2p$ angles that make the objective function large. Again we will discuss how this is achieved in section \ref{from_small_to_large}. But once the parameters are fixed we see that each tossed graph has nearly the same value of the objective function since the standard deviation over the 25 samples is small.

\begin{table}[htbp]
\begin{mdframed}[
    tikzsetting={align=center,draw=black,ultra thick,align=center},
    innerrightmargin=5pt,innerleftmargin=5pt,innerbottommargin=5pt
    ]

\centering
\setlength{\tabcolsep}{20pt}
\vspace{2mm}
\begin{tabular}{c|cc|cc|cc} & \multicolumn{2}{c|}{Low} & \multicolumn{2}{c|}{Random} & \multicolumn{2}{c}{High} \\
$p$ & Mean & Std. & Mean & Std. & Mean & Std. \\
\hline
2 & 6.636 & 0.319 & 14.691 & 0.036 & 22.409 & 0.228 \\
3 & 5.218 & 0.294 & 15.125 & 0.042 & 23.109 & 0.175 \\
4 & 3.933 & 0.259 & 14.627 & 0.157 & 23.822 & 0.272 \\
5 & 3.132 & 0.159 & 15.725 & 0.113 & 24.349 & 0.179 \\
6 & 2.550 & 0.100 & 16.404 & 0.140 & 24.918 & 0.266 \\
7 & 1.954 & 0.088 & 15.975 & 0.096 & 25.110 & 0.221 \\
\end{tabular}
\caption{Mean and standard deviation of objective function values across 25 random 20-node-3-regular graph instances each having a MaxCut value of 26. For each $p$ there are 3 sets of fixed parameters. The first is chosen to give a low cost function value, the second is randomly chosen and the third is chosen to make the cost function high. In the low and high cases the angles are not the best possible, but are chosen to drive the objective function away from a random point in the landscape.}
\end{mdframed}
\end{table}
So far we have an argument that predicts concentration for low $p$. But the chart shows that concentration is maintained when $p$ increases. We discuss this next.


\section{Higher $\protect\scalebox{1.25}{$\mathit{p}$}$}\label{higher_p_section}

Return to Eq. (\ref{objective_big}) and consider one term associated with edge $\langle jk \rangle$.  The middle term is conjugated by $p$ operators of the form $U(C,\gamma)$ as in Eq. (\ref{driver}).  Each conjugation may bring in new qubits which can be thought of as being reached by taking $p$ steps from $j$ and $k$.  If there are no backward steps and only new qubits are reached we can picture the qubits connected to $j$ or $k$ as vertices on a tree. For example for $p=1$ we have the tree seen in the third picture in equation (\ref{subgraph_types}).  This has 6 qubits. If we go to $p=2$ the tree has 14 qubits. For $p$ ranging from 2 to 7 the relevant tree sizes are 14, 30, 62, 126, 254 and 510.  So working with 20 qubit graphs, we see that certainly by $p=3$ the qubits connected to each edge are covered many times by the conjugation operation.  We have left the domain of fixed $p$ with $n$ large.  However if you look at the data in Table 1 we see that even for $p=7$ we have concentration. Thus we need another explanation for the concentration, other than what was shown in Section \ref{fixed_p_section}.

We are going to make use of the fact that the objective function we evaluate is a sum of $m$ individual terms,
\begin{equation}\label{objective_individual_terms}
    F(\boldsymbol{\gamma}, \boldsymbol{\beta}) = \sum_{\alpha=1}^{m} \bra{\boldsymbol{\gamma}, \boldsymbol{\beta}} C_{\alpha} \ket{\boldsymbol{\gamma}, \boldsymbol{\beta}} \text{ .}
\end{equation}
We always think of the parameters $\boldsymbol{\gamma}$ and $\boldsymbol{\beta}$ as fixed. The randomness comes from tossing instances. So for the rest of this section we stop carrying around the $\boldsymbol{\gamma}$ and $\boldsymbol{\beta}$ and we write
\begin{equation}\label{stop_caring}
    F = \sum_{\alpha = 1}^{m} \mathcal{C_{\alpha}} \text{ .}
\end{equation}
We have some distribution of $n$-bit instances of some combinatorial search problem.  We can think of each term in (\ref{stop_caring}) as a random variable whose value lies between 0 and 1.  The sum is of order the number of clauses $m$.  In most cases of interest $m$ is linear or quadratic in $n$.  By the Law of Large Numbers, if the terms in (\ref{stop_caring}) are independent then the standard deviation of $F$, regardless of the distribution, is of order $\sqrt{m}$.  In this case we have concentration as typical instances will have the same value of $F$ up to corrections of order $\sqrt{m}$.  So we need to investigate the correlations between individual terms in Eq. (\ref{stop_caring}).  We do this by example, returning again to MaxCut on 3-regular graphs.

For MaxCut each clause is labeled by an edge so we can think of $\alpha$ as an edge label.  In (\ref{stop_caring}) the individual terms are correlated because they correspond to edges on the same graph.  Let us see if we can use our simulations to estimate the correlations between different edges. First we need to be explicit about the random process that generates $\mathcal{C}_{\alpha}$. Pick a random $n$-bit-3-regular graph. Toss a random permutation of the integers $1,2,\dots,m$ and use this sequence to relabel the edges. (We randomly relabel the edges so that the edge name has no possible connection with the process that generated the graph.) Take the quantum expectation of the clause corresponding to edge $\alpha$ in the quantum state determined by the graph. (Recall again that the angles are fixed.) This defines random variables that we call $\mathcal{C}_{\alpha}$, $\alpha = 1, \ldots, m$. 

Now any $\mathcal{C_{\alpha}}$ has the same distribution as $\mathcal{C}_{1}$. So we have
\begin{equation}\label{independence_equality}
    \E[{\mathcal{C}}_{\alpha}] = \E[{\mathcal{C}}_{1}] \quad\text{and}\quad \std(\mathcal{C}_{\alpha}) = \std(\mathcal{C}_{1}) \text{ .}
\end{equation}
The correlation between $\mathcal{C_{\alpha}}$ and $\mathcal{C_{\alpha^{\prime}}}$ is
\begin{equation}\label{correlation_coef}
    \Corr(\mathcal{C_{\alpha}},\mathcal{C_{\alpha^{\prime}}}) = \frac{\Cov(\mathcal{C_{\alpha}},\mathcal{C_{\alpha^{\prime}}})}{\std(\mathcal{C_{\alpha}}) \std(\mathcal{C_{\alpha^{\prime}}})}
\end{equation}
where
\begin{gather}
    \Cov(\mathcal{C_{\alpha}}, \mathcal{C_{\alpha^{\prime}}}) = \E \big[ (\mathcal{C_{\alpha}} - \E[\mathcal{C_{\alpha}}])(\mathcal{C_{\alpha^{\prime}}} - \E[\mathcal{C_{\alpha^{\prime}}}])\big],
\end{gather}
where the expectation is taken over the random graph and the random permutation, and the standard deviation is the square root of the variance
\begin{equation}
    \var(\mathcal{C_{\alpha}}) = \E \big[(\mathcal{C_{\alpha}} - \E[\mathcal{C_{\alpha}}])^{2} \big] \text{ .}
\end{equation}
Using (\ref{independence_equality}) and (\ref{correlation_coef}) the correlation coefficient between $\mathcal{C_{\alpha}}$ and $\mathcal{C_{\alpha^{\prime}}}$ can be written as
\begin{equation}
    \Corr(\mathcal{C_{\alpha}}, \mathcal{C_{\alpha^{\prime}}}) = \frac{\Cov(\mathcal{C_{\alpha}, \mathcal{C_{\alpha^{\prime}}})}}{\var(\mathcal{C}_{1})} \text{ .}
\end{equation}
Now consider the variance of $F$
\begin{equation}
    \var(F) = \E \big[(F - \E[F])^{2} \big] \text{ .}
\end{equation}
Since $F$ is a sum, see (\ref{stop_caring}), we have that
\begin{equation}
    \var(F) = \sum_{\alpha = 1}^{m} \var(\mathcal{C_{\alpha}}) + \sum_{\alpha \neq \alpha^{\prime}} \Cov(\mathcal{C_{\alpha}}, \mathcal{C_{\alpha^{\prime}}})
\end{equation}
so
\begin{equation}
    \sum_{\alpha \neq \alpha^{\prime}} \Cov(\mathcal{C_{\alpha}}, \mathcal{C_{\alpha^{\prime}}}) = \var(F) - \sum_{\alpha = 1}^{m} \var(\mathcal{C_{\alpha}}) \text{ .}
\end{equation}
Inside each graph the $C_{\alpha}$ are correlated. The random permutation guarantees that the correlation is the same for each pair $C_{\alpha}$, $C_{\alpha^{\prime}}$.  So each covariance is 
\begin{equation}
    \frac{1}{m(m - 1)} \big[ \var(F) - m \var(\mathcal{C}_{1}) \big]
\end{equation}
and each correlation coefficient is
\begin{equation}\label{correlation_coef_est}
    \frac{1}{m - 1} \Bigg[\frac{\var(F)}{m \var(\mathcal{C}_{1})} - 1 \Bigg] \text{ .}
\end{equation}
We want to toss random graphs to estimate this quantity. Suppose we toss $N$ graphs (we will use $N=100$). For fixed angles, for each of the $N$ graphs we calculate $F$ via (\ref{objective_individual_terms}) and get $N$ samples $\mathcal{F}_{k}$ with $k=1,\dots, N$, where the mean is estimated as
\begin{equation}
    \overline{F} = \frac{1}{N} \sum_{k=1}^{N} \mathcal{F}_{k}
\end{equation}
and the variance is estimated as
\begin{equation}\label{var_est}
    \frac{1}{N-1} \sum_{k=1}^{N} (\mathcal{F}_{k} - \overline{F})^{2} \text{ .}
\end{equation}
Next we need to estimate $\var(\mathcal{C}_{1})$. Since the mean and variance of each $\mathcal{C}_{\alpha}$ are the same as for $\mathcal{C}_{1}$, we want to use all $mN$ samples for the estimate.
We can write for the estimate of $\E[\mathcal{C}_{1}]$
\begin{equation}
\begin{split}
    \overline{\mathcal{C}} &= \frac{1}{mN}\sum_{\alpha=1}^{m} \sum_{k=1}^{N} \mathcal{C}_{\alpha, k} \\
                              &= \frac{1}{m} \overline{F}
\end{split}
\end{equation}
and for the estimate of $\var(\mathcal{C}_{1})$ we use
\begin{equation}\label{big_var2}
 \frac{1}{mN - 1} \sum_{\alpha=1}^{m} \sum_{k=1}^{N} (\mathcal{C}_{\alpha, k} - \overline{\mathcal{C}})^{2} \text{ ,}
\end{equation}
where the $mN-1$ makes the estimate unbiased (in the uncorrelated case).
We use (\ref{big_var2}) to estimate $\var(\mathcal{C}_{1})$ and (\ref{var_est}) to estimate $\var(F)$ and obtain our estimate of the correlation coefficient via (\ref{correlation_coef_est}).

Working at 20 bits we tossed random 3-regular graphs with the MaxCut value fixed at 26 to numerically investigate the correlation coefficient given by (28).  Here we worked at p=8 with five sets of fixed parameters.  One set was chosen to produce a low value of the objective function, one set at medium low value, one set at a medium high value and one set at a high value. (Presumably, fixing the MaxCut value reduces the spread of the objective function especially when using the high value parameters.) We also included a set of random parameters. We tossed 100 graphs for each set of parameters.  The results are shown in Table 2.  In general the correlation coefficient (20) can range from -1 which is perfectly anti-correlated to 1 which is perfectly correlated. (In our case the correlation coefficient must be $\geq \frac{-1}{m-1}$ as can be seen in (\ref{correlation_coef_est}).)  The observed values we get near zero are good evidence that the individual terms that sum to make $F$ have tiny correlations. Working at 20 bits this can help explain why the objective function is nearly independent of the tossed graph. If we go to higher bit number we imagine that the correlation coefficient will tend to zero because in bigger graphs the edges are further apart.

\begin{table}[htbp]

\begin{mdframed}[
    tikzsetting={align=center,draw=black,ultra thick,align=center},
    innerrightmargin=5pt,innerleftmargin=5pt,innerbottommargin=5pt
]

\centering
\setlength{\tabcolsep}{20pt}
\vspace{2mm}
\begin{tabular}{c|c|c|c} & $\overline{F}$ & $\widehat{\std}(F)$ & Correlation Coefficient \\
\hline
Low & 2.311 & 0.118 & -0.016 \\
Med Low & 7.621 & 0.297 & 0.013 \\
Random & 15.203 & 0.262 & 0.040 \\
Med High & 21.4978 & 0.250 & 0.008 \\
High & 24.995 & 0.220 & -0.031 \\
\end{tabular}
\caption{Each row has a set of fixed parameters at $p=8$. The data comes from tossing 100 3-regular graphs with a MaxCut of 26. In the first row the parameters are chosen (see section \ref{from_small_to_large}) to produce a low value of the cost function. In the second row the parameters are chosen for a medium low value. Etc. In all cases the correlation coefficient is small.}
\end{mdframed}
\end{table}


\section{From small to large instances}\label{from_small_to_large}

We have observed in our numerical experiments that for any $p$, if we fix parameters such that the objective function has a high value at some small number of qubits then those same parameters will produce a high value at a larger number of qubits. When we looked for very good parameters so that the objective function value is close to optimal at low bit number then these same parameters produced a near optimal value of the objective function at a higher bit number. We illustrate this with an example.  We tossed a random 3-regular graph at 10 bits and worked at $p=8$.  We searched for good parameters by doing 200 random restarts with a canned Matlab optimization algorithm running on a laptop.  This produced a set of angles with an approximation ratio of 0.984.  It is possible that further searching would have led to an even higher approximation ratio but our goal was not to find the very best parameters.  We then used these 16 angles and evaluated the cost function on 25 randomly chosen 3-regular graphs with 24 vertices.  The average approximation ratio for the 25 instances was 0.934 with a standard deviation of 0.014.  This value for the approximation ratio is high and it came without any searching at 24 qubits.

The good parameters we found at 10 bits at $p=8$ will not work well on arbitrarily large graphs. If the number of vertices is very large, then working at $p=8$, the local pictures on a very large graph will not cover the graph. For $p=8$, working on very big 3-regular graphs the largest number of qubits that can be involved in any clause is 1022.  If we have a graph with tens of thousands of vertices then these local terms will not see the large loops. And it is only odd length loops that prevent a MaxCut instance from being fully satisfiable. The QAOA will not be able to tell the difference between a graph with large loops that are all even length from a graph with large odd length loops. So certainly in worst case the QAOA will not work as well on large graphs with fixed $p$ and parameters found on small graphs.

In the near future, experimentalists will provide us with gate model quantum computers with up to 100 qubits. Suppose we run the QAOA on hardware to find good approximation ratios for MaxCut on 3-regular graphs that are randomly generated. Then certainly at $p=8$ we are well out of the fixed $p$ large $n$ regime.  It may be that we can find good parameters with a classical simulator at say 20 bits and these same angles will produce good approximation ratios when run on a quantum computer at high bit number. This is a way to test the QAOA and the hardware in a regime that can not be classically simulated.

\section{Future Outlook}

The main finding of this paper is that if we look at the objective function of the QAOA (\ref{objective_f}) then for fixed $\boldsymbol{\gamma}$ and $\boldsymbol{\beta}$ this function has the same value on all typical instances generated by some reasonable distribution. For fixed $p$ and search problems where the number of clauses in which any variable can appear is bounded, we can prove that for $n$ large there is concentration with fluctuations of order the square root of the number of clauses.  For $p$ growing with $n$, we can argue for the concentration using the Law of Large Numbers and the assumption that individual terms in the objective function are not very correlated.  Our numerical experiments bolster these arguments.

This leads to a strategy for running the QAOA on a quantum computer where the task is to find a good approximation ratio for some combinatorial search problem.  Suppose we have a guarantee that the instances come from some reasonable distribution.  Then given the first instance to solve, work hard to find the optimal parameters. This may be done by conducting a variational search for good parameters making repeated calls to the quantum computer.  Or we may adopt a leapfrogging strategy as will be discussed in the next paragraph. However once good parameters have been found for one instance, these same parameters can be used on other typical instances. It might be possible to improve performance for each instance by doing a narrow search in the neighborhood of the good parameters.  Whether this is cost effective will depend on the problem at hand, the expense of running the quantum device and the desired performance. However our findings eliminate the need to blindly search on each instance and gets rid of (or significantly lessens)  the burden of the outer loop optimization beyond the first instance.

Another key finding is that if we fix $p$, then parameters that bring the cost function to a high value at low bit number will also bring the cost function to a high value on larger instances. In our numerical simulations we found good parameters at 10 bits and using these parameters we got good performance at 20 and 24 bits. There was no additional optimization at the higher bit numbers.  This suggests a strategy for finding good parameters.  Say we have a quantum computer with 100 qubits and we are looking at random instances of MaxCut on some distribution of graphs.  Toss a random instance at say 20 bits and use a classical computer to find optimal angles.  Here $p$ is fixed at say 6.  Once the 12 angles are found we could run the quantum computer on a randomly chosen instance at 50 qubits using these 12 parameters.  Refine the angles by doing a local search near the given angles. This may take some time on the quantum computer but the starting point should already be good.  Now we have a new set of 12 parameters which are working well at 50 qubits.  Use these same parameters on randomly chosen instances as 100 bits.  Further refine if desired. This leapfrogging strategy will greatly reduce the computational cost of the outer loop optimization.

We have introduced strategies for running the Quantum Approximate Optimization Algorithm that greatly reduce the number of function calls to the quantum computer compared with direct variational approaches.  This may shorten the time it will take until we can run the QAOA on a near term device and test its performance in regimes where classical simulation is not available. We look forward to these experiments.

\section{Acknowledgements}
We thank the Google AI Quantum team for useful discussion. EF also thanks Soonwon Choi, Misha Lukin, Hannes Pichler, Sheng-Tao Wang and Leo Zhou for many good chats. We acknowledge Jeffrey Goldstone for help with the acknowledgements. The work of EF was partially supported from NSF grant CCF-1729369 and ARO contract W911NF-17-1-0433. FB work is partially supported by NSF.

\end{document}